\documentclass[11pt]{article}
\usepackage{amssymb,amsmath}
\usepackage{latexsym}
\usepackage{color}
\usepackage{epsfig}
\usepackage{tensor}
\newcommand\p\partial

\newcommand{\bea}{\begin{eqnarray}}
\newcommand{\eea}{\end{eqnarray}}
\newcommand{\be}{\begin{equation}}
\newcommand{\ee}{\end{equation}}

\begin{document}
\vskip 1in

\begin{center}
{\Large \bf Topologically Massive Gravity from the Outside In}
 \vspace*{0.5cm}\\
 {Colin Cunliff$^{\flat}$
 }
\end{center}
 \vspace*{0.1cm}

 \begin{center}
 $^{\flat}${\it Department of Physics,
 University of California, Davis, \\
 Davis, CA 95616, USA\\
  {\tt cunliff@physics.ucdavis.edu}
 }
 \vspace{0.15cm}

 \end{center}

 \vspace{0.15cm}

 \begin{abstract}

The asymptotically anti-de Sitter solutions of cosmological
topologically massive gravity (TMG) are analyzed for values of the
mass parameter in the range $\mu\geq1$.  At non-chiral values, a new
term in the Fefferman-Graham expansion is needed to capture the bulk
degree of freedom.  The CDWW modes provide a basis for the pure
non-Einstein solutions at all $\mu$, with nonlinear corrections
appearing at higher order in the expansion.

 \vspace{12pt}
 Pacs:  04.20.-q,04.60.-m,04.70.-s,11.30.-j
 \vspace{12pt}
 \end{abstract}

Topologically massive gravity (TMG)\cite{Deser1981wh,Deser1982vy}
with a negative cosmological constant appears to be the simplest
theory that contains both black holes and local gravitational
degrees of freedom, making it a potentially useful toy model to
explore various questions in quantum gravity.  However, at arbitrary
values of the coupling constants, the theory appears to be unstable,
containing either positive mass black holes and negative energy
gravitons, or the reverse, depending on the choice of sign of the
Einstein-Hilbert piece of the action. Recently, Li et al.\
\cite{Li2008dq} proposed that with suitable boundary conditions, the
local bulk degree of freedom disappears at the ``chiral'' coupling
$\mu \ell =1$, allowing the possibility of choosing the sign of the
action such that only positive mass black holes are included. With
these boundary conditions, the theory becomes chiral, in that the
asymptotic symmetry consists of a single copy of the Virasoro
algebra \cite{Strominger2008dp,Carlip2008qh,Henneaux2009pw}, and the
theory was dubbed chiral gravity \cite{Li2008dq,Maloney2009ck}.

The claim that chiral gravity admits only non-negative energy modes
has been the subject of much debate in the literature. Several
non-perturbative studies found a single local propagating degree of
freedom at all values of $\mu$
\cite{Carlip2008qh,Grumiller2008,Blagojevic2008}; however, the
boundary conditions satisfied by these modes was not investigated.
At the linearized level, other authors found negative-energy modes
at $\mu \ell = 1$
\cite{Carlip2008jk,Carlip2008eq,Grumiller2008qz,Giribet2008bw}, but
these modes either were not chiral or required different boundary
conditions than those used in \cite{Li2008dq}. After some confusion
in the literature, Maloney et al.\ \cite{Maloney2009ck} concluded
TMG at the critical value could be divided into two theories
depending on the choice of boundary conditions:  chiral gravity with
Brown-Henneaux \cite{Brown1986nw} boundary conditions and log
gravity with relaxed boundary conditions that include a logarithmic
term in the asymptotic expansion of the metric.  For related work on
boundary conditions, see \cite{Henneaux2009pw,Grumiller2008es}.
Additionally it was shown in \cite{Maloney2009ck} that all
stationary, axially symmetric solutions of chiral gravity are the
familiar BTZ black holes \cite{Banados1992gq} and have non-negative
energy. The authors found that the proposed counterexamples either
required the relaxed boundary conditions, and thus were solutions to
log gravity, or developed linearization instabilities at second
order, and they speculated that all asymptotically anti-de Sitter
non-Einstein solutions at the critical point are in fact solutions
to log gravity.

Recently, Comp\`{e}re et al.\ \cite{Compere2010} discovered a new
class of non-Einstein solutions of chiral gravity using the
Fefferman-Graham expansion \cite{Fefferman1985ab} with
Brown-Henneaux boundary conditions. They further examined a subset
of the general solution that included linear perturbations from
AdS$_3$ and BTZ backgrounds. Of the solutions examined, all
contained either naked singularities or closed timelike curves and,
unless they can be excluded as unphysical, may render chiral gravity
unstable.

This paper extends their work to non-chiral values $\mu \ell
>1$. New terms in the Fefferman-Graham expansion are needed to
capture the bulk degrees of freedom at all values of the mass
parameter, and for each value of $\mu$, a similar phenomenon occurs:
one of the equations of motion disappears, leaving one piece of the
metric unconstrained by the equations of motion. Section 2 gives the
solution to second order in this new term. The division between
Einstein and non-Einstein solutions becomes explicit in this
formalism: one set of terms in the expansion captures all Einstein
solutions, and the second set captures the non-Einstein solutions.
Section 3 examines the special case of chiral gravity, where the
solution of \cite{Compere2010} is given in light-cone coordinates.
In this formalism, we see the chiral point is just the point at
which the two sets of terms overlap.  Section 4 maps the CDWW modes
\cite{Carlip2008jk} onto the full solution of Section 1. The CDWW
modes provide a complete basis of the non-Einstein solutions only,
and additional ingredients are needed to include the Einstein
solutions.  The solution given in Section 2 agrees with CDWW to
second order; however, nonlinear deviations from CDWW are found for
several integral values of $\mu$ at higher order and are likely to
exist for generic $\mu$. I conclude with a discussion on the
significance of these solutions on the stability of chiral gravity.

\section{Asymptotic solution of TMG}
The equation of motion for topologically massive gravity (TMG)
\cite{Deser1981wh,Deser1982vy} with a cosmological constant is
\begin{equation}
  R_{\mu \nu} - \frac{1}{2}Rg_{\mu \nu} + \Lambda g_{\mu \nu} +
  \frac{1}{\mu} C_{\mu \nu} = 0, \label{eom1}
\end{equation}
where $C_{\mu \nu}$ is the Cotton tensor
\begin{equation}
  C\indices{_{\mu \nu}} = \epsilon\indices{_\mu^{\alpha \beta}}
  \nabla\indices{_\alpha} \left(R\indices{_{\beta \nu}} -
  \frac{1}{4}Rg\indices{_{\beta \nu}} \right). \label{cotton}
\end{equation}
In the discussion below, we work in units where $\Lambda = -1$.  By
the Bianchi identity, the Cotton tensor is symmetric, traceless, and
covariantly conserved.  Taking the trace of \eqref{eom1}, we find
the solution is a spacetime of constant scalar curvature
\begin{equation}
  R=-6,
\end{equation}
and the equation becomes
\begin{equation}
  R_{\mu \nu} + 2g_{\mu \nu} + \frac{1}{\mu} C_{\mu \nu}=0.
  \label{eom2}
\end{equation}
Pure Einstein solutions are those for which $R_{\mu \nu} = -2g_{\mu
\nu}$, and it is apparent from \eqref{cotton} that the Cotton tensor
is identically zero for all Einstein metrics.  Thus TMG contains all
of the ordinary Einstein solutions plus the massive propagating
modes for which the Cotton tensor is non-zero.

In particular, TMG admits an asymptotically anti-de Sitter (AdS)
solution which can be written in Gaussian normal coordinates as
\begin{equation}
  ds\indices{^2} = dr\indices{^2} + g\indices{_{ij}}dx\indices{^i}
  dx\indices{^j}, \label{gausscoord}
\end{equation}
where
\begin{equation}
  g\indices*{^{(0)}_{ij}} = \lim_{r \to \infty} e\indices{^{-2r}}g_{ij}(x,r)
\end{equation}
is the metric on the boundary.  Here we take Greek indices to run
over all coordinates, and Latin indices run over the non-radial
coordinates. Due to a theorem by Fefferman and Graham
\cite{Fefferman1985ab}, the two-dimensional metric $g_{ij}$ can
always be expanded near the boundary in powers of $e^r$:
\begin{equation}
  g\indices{_{ij}} = e\indices{^{2r}}\left( g\indices*{^{(0)}_{ij}} +
  e\indices{^{-2r}}g\indices*{^{(2)}_{ij}} + e\indices{^{-4r}}
  g\indices*{^{(4)}_{ij}} + \cdots\right). \label{FGEins}
\end{equation}
For three-dimensional Einstein gravity, the expansion terminates at
$g_{(4)}$ (all higher order terms are zero), and the first three
terms are sufficient to capture all AdS$_3$ and BTZ solutions to
pure
Einstein gravity \cite{Skenderis1999nb}.  

As noted in \cite{Skenderis2009nt}, the form of the expansion
depends on the bulk theory, and other theories may exhibit different
asymptotic behavior.  Because all solutions of Einstein gravity are
also solutions of TMG, we need at least the terms in \eqref{FGEins}
in the expansion for TMG. However, these terms alone do not capture
the propagating degree of freedom of TMG: at $\mu\neq 1$, the series
\eqref{FGEins} still terminates at $g_{(4)}$, and the
Cotton tensor vanishes to all orders. 
Thus to include all Einstein and non-Einstein solutions, at all
values of the mass parameter $\mu>1$, the expansion is\footnote{Note
that I propose this new series as an \emph{ansatz}. This differs
from the view of \cite{Skenderis2009nt} in which the correct
asymptotic expansion should be \emph{derived} from the bulk theory
via the AdS/CFT correspondence.  However, this should not affect the
outcome -- the solutions found using this expansion still satisfy
the equations of motion.}
\begin{eqnarray}
  g\indices{_{ij}} &=& e\indices{^{2r}}\left(g\indices*{^{(0)}_{ij}} +
    e\indices{^{-2r}}g\indices*{^{(2)}_{ij}} +
    e\indices{^{-4r}}g\indices*{^{(4)}_{ij}}\right) \quad \quad \quad
    \quad \quad \quad \quad \quad
    \text{Einstein} \nonumber \\
  &+& e\indices{^{2r}}\left( e\indices{^{-(\mu + 1)r}}g\indices*{^{(\mu + 1)}_{ij}} +
e\indices{^{-(\mu+3)r}}g\indices*{^{(\mu+3)}_{ij}} + \cdots \right)
\quad \quad \quad \text{non-Einstein} \label{FGTMG}
\end{eqnarray}
In this form, it becomes apparent that the odd integral values of
the mass parameter are just the values at which the two expansions
overlap.

Now the procedure is to plug the expansion into the equations of
motion \eqref{eom2}, collect terms of the same power of $e^r$, and
set the coefficients equal to zero.  Thus we can solve for the
higher-order terms \emph{algebraically} in terms of the lower order
terms.  The boundary metric $g_{(0)}$ may be left as a free field;
however, the process is simplified by the use of a constant boundary
metric and light-cone coordinates $x^+=\frac{1}{\sqrt{2}}(t+\phi)$
and $x^-=\frac{1}{\sqrt{2}}(t-\phi)$.  Consistent with the
conventions of \cite{Compere2010}, I set $g^{(0)}_{+-}=-1$ and
$\sqrt{-g}\epsilon^{r-+}=1$.  The solutions for $g_{(2)}$ and
$g_{(4)}$ at non-critical values are
\begin{eqnarray}
  g\indices*{^{(2)}_{++}} = L(x^+)  &\text{and}& g\indices*{^{(2)}_{--}} = \bar{L}(x^-) \nonumber\\
  &g\indices*{^{(4)}_{+-}} = -\frac{1}{4} L(x^+)\bar{L}(x^-). &
  \label{Esoln}
\end{eqnarray}
As noted previously, the solutions encompassed in these terms
contain \emph{only} the ordinary Einstein solutions
\cite{Skenderis1999nb}, including the BTZ black hole
\cite{Banados1992gq}.  For $\mu>1$, the equations set
$g^{(\mu+1)}_{+-}$ and $g^{(\mu+1)}_{--}$ to zero, but the equations
for $g^{(\mu+1)}_{++}$ disappear, and this component is
unconstrained by the theory.  The solution is
\begin{eqnarray}
  g\indices*{^{(\mu+1)}_{++}} &=& F(x^+,x^-) \label{soln}\\
  g\indices*{^{(\mu+3)}_{++}} &=&
      \frac{1}{2\mu+6} \partial_+ \partial_- F \nonumber\\
  g\indices*{^{(\mu+3)}_{+-}} &=& -\frac{(\mu+1)^2-2}{2\mu(\mu+3)}
      \bar{L}F -
      \frac{1}{2\mu(\mu+3)}
      \partial\indices*{_-^2}F \nonumber\\
  g\indices*{^{(\mu+5)}_{++}} &=& \cdots \nonumber .
\end{eqnarray}
The calculation for generic $\mu$ was performed by hand.
Additionally, a Maple script was written to solve the equations of
motion order-by-order at integral values of $\mu$, and results match
\eqref{soln} for $\mu=1$, 2 and 3.  For the full solution
\eqref{Esoln} and \eqref{soln}, we find a non-zero Cotton tensor
(and therefore non-Einstein solutions) \emph{only} when $F(x^+,x^-)
\neq 0$.  Thus, the function $F$ contains the massive, propagating
modes unique to TMG.

\section{Chiral Gravity}
Two new features appear at the critical point $\mu=1$.  First, the
asymptotic expansion acquires a logarithmic term
\cite{Skenderis2009nt}
\begin{equation}
  g\indices{_{ij}} = e\indices{^{2r}}\left( g\indices*{^{(0)}_{ij}}
  + re\indices{^{-2r}}g\indices*{^{(1)}_{ij}} + e\indices{^{-2r}}g\indices*{^{(2)}_{ij}} + \cdots
  \right) \label{FGchiral}.
\end{equation}
As discovered by Grumiller and Johansson \cite{Grumiller2008qz}, the
logarithmic term contains a new branch of solutions only accessible
at $\mu=1$.  At $\mu \neq 1$ the equations of motion force
$g_{(1)}=0$, and this term is not present in the general expansion
\eqref{FGTMG}. Chiral gravity \cite{Li2008dq,Maloney2009ck} is the
subset of the theory at $\mu=1$ with the logarithmic mode turned
off.  Here the functions $F(x^+,x^-)$ and $L(x^+)$ overlap, and the
general solution in light-cone coordinates to third order
($g_{(6)}$) is
\begin{eqnarray}
  g^{(2)}_{++} &=& F(x^+,x^-)\nonumber \\
  g^{(2)}_{--} &=& \bar{L}(x^-)\nonumber\\
  g^{(4)}_{++} &=& \frac{1}{8} \partial_+ \partial_- F \nonumber\\
  g^{(4)}_{+-} &=& -\frac{1}{4} \bar{L} F -
      \frac{1}{8} \partial_-^2 F \nonumber\\
  g^{(6)}_{++} &=& \frac{1}{8} F\partial_-^2 F +
      \frac{1}{96}\partial_+^2 \partial_-^2 F - \frac{1}{96}
      \left(\partial_-F\right)^2 \nonumber\\
  g^{(6)}_{+-} &=& -\frac{1}{16}\bar{L}
      \partial_+ \partial_- F - \frac{1}{96} \partial_+ \partial_-^3 F \nonumber\\
  g^{(6)}_{--} &=& \frac{1}{9} \bar{L}
      \partial_-^2 F + \frac{1}{144} \left(\partial_-
      \bar{L}\right)\left(\partial_- F\right) + \frac{1}{288}
      \partial_-^4 F \label{chiralsoln}
\end{eqnarray}
In agreement with \cite{Compere2010}, we find that Einstein
solutions are the subset for which the function $F$ depends only on
$x^+$. When $\partial_-F=0$, the Cotton tensor vanishes to all
orders, and the expansion \eqref{FGchiral} terminates at $g_{(4)}$.
Note that the requirement for vanishing Cotton tensor at the chiral
value is more stringent that at non-chiral values, for which $F=0
\leftrightarrow C_{\mu \nu} = 0$.

\section{Revisiting CDWW}
Previously, Carlip et al.\ \cite{Carlip2008jk,Carlip2008eq} found a
complete set of solutions -- the CDWW modes -- to the linearized
equations of motion at all values of $\mu$.  These solutions share
some important features with \eqref{soln}, namely
\begin{itemize}
  \item the solutions are invariant under $\mu \rightarrow -\mu$ and
  a chirality flip ($x^+ \leftrightarrow x^-$), and
  \item they exhibit a $\mu$-dependent asymptotic behavior, with
  different fall-off conditions for each component of the metric.
\end{itemize}
Given the similarities, it's natural to ask if \eqref{soln} contains
the CDWW modes.

However, CDWW solved for the linearized Einstein tensor, and these
modes must first be converted into perturbations of the metric in
Gaussian normal coordinates before a direct comparison can be made.
This can always be done in three dimensions, since a perturbation of
the Einstein tensor uniquely determines a perturbation in the
metric.  The solutions for each component of the linearized Einstein
tensor are the Bessel functions given in eqn. (A.28) of
\cite{Carlip2008eq}, for example
\begin{equation}
  \mathcal{H}_{++} = \frac{\omega_+^2}{\omega} e^{i[\omega_+x^+ +
  \omega_-x^-]}zJ_{\mu-2}(\omega z) + h.c. \label{CDWW}
\end{equation}
is shown here for comparison.  In this expression, $z$ is the radial
coordinate related to the $r$ of the previous section by $z=e^{-r}$;
$\omega_+$ and $\omega_-$ are eigenvalues of the $SL(2, \mathbb{R})$
generators $i\partial_+$ and $i\partial_-$, and $\omega^2=-2\omega_+
\omega_-$. Eqn (5.5) of \cite{Carlip2008eq} relates the Einstein
tensor to the metric perturbations via differential equations such
as
\begin{equation}
  \mathcal{H}_{++} = -\frac{1}{2}z\partial_z (z\partial_z + 2)
  g_{++}. \label{gauss}
\end{equation}
The final step is to expand the modes \eqref{CDWW} in powers of $z$
and solve \eqref{gauss} for the metric perturbations order by order.
The CDWW modes as \emph{metric peturbations} are
\begin{eqnarray}
  g\indices{_{++}} &=& \frac{-\omega_+^2
      \omega^{\mu-3}}{2^{\mu-3}(\mu+1)\Gamma(\mu)}
      e\indices{^{i[\omega_+x^+ + \omega_-x^-]}}\left[ z^{\mu-1} +
      \frac{\omega^2 z^{\mu+1}}{2^2(\mu+3)} +
      \frac{\omega^4z^{\mu+3}}{2^5(\mu+5)(\mu+3)\mu}\right]
      \nonumber \\
  g\indices{_{+-}} &=& \frac{\omega^{\mu+1}}{2^\mu(\mu+3)\Gamma(\mu+2)}e^{i[\omega_+x^+ + \omega_-x^-]}\left[z^{\mu+1}
      -\frac{\omega^2}{2^2(\mu+5)}z^{\mu+3}\right]\nonumber \\
  g\indices{_{--}} &=& \frac{-\omega_-^2 \omega^{\mu + 1}}{2^{\mu +
      1}(\mu+5)\Gamma(\mu + 4)} e\indices{^{i[\omega_+x^+ +
      \omega_-x^-]}} z\indices{^{\mu+3}} \label{CDWWmetric}
\end{eqnarray}
With the identification
\begin{eqnarray}
  F(x^+,x^-) = -\frac{\omega_+^2
  \omega\indices{^{\mu-3}}}{(\mu+1)2^{\mu-3}\Gamma(\mu)}
  e^{i[\omega_+x^+ + \omega_-x^-]} \label{CDWWmode}
\end{eqnarray}
the CDWW modes can be written as
\begin{eqnarray}
  g\indices{_{++}} &=& F(x^+,x^-)z^{\mu-1} +
      \frac{z^{\mu+1}}{2(\mu+3)}\partial_+ \partial_- F +
      \frac{(\mu+1)z^{\mu+3}}{8(\mu+5)(\mu+3)\mu}\partial_+^2
      \partial_-^2 F
      \nonumber \\
  g\indices{_{+-}} &=& \frac{-z^{\mu+1}}{2\mu(\mu+3)}\partial_-^2 F -
      \frac{z^{\mu+3}}{4(\mu+5)(\mu+3)\mu} \partial_+ \partial_-^3 F \nonumber \\
  g\indices{_{--}} &=&
  \frac{z\indices{^{\mu+3}}}{4(\mu+5)(\mu+3)(\mu+2)\mu}\partial_-^4
  F
\end{eqnarray}
Comparison with the asymptotic solutions \eqref{soln} and
\eqref{chiralsoln} reveals that CDWW correctly gives all numerical
coefficients for pure derivatives of $F(x^+, x^-)$.  However, the
full solution is more than just a background Einstein solution plus
the CDWW modes. It also contains interactions between the background
and CDWW, e.g.\ the piece proportional to $\bar{L}F$ in
\eqref{soln}.  Additionally, nonlinear deviations from CDWW appear
at higher orders.  In the chiral solution \eqref{chiralsoln}, for
example, CDWW fails to include the non-linear terms $F\partial_-^2F$
and $(\partial_-F)^2$ in the $g^{(6)}_{++}$ component.

While the $(\mu+5)$ and higher order solution at generic $\mu$ has
not been found, solutions at integral values of $\mu$ have been
explored using Maple, and several patterns emerge.  For all $\mu$
examined, the first nonlinear deviation from CDWW appears in the
$g_{++}$ component at order $2\mu+4$ and contains terms proportional
to $F\partial^2_-F$ and $\left(\partial_-F\right)^2$.  In the case
of odd $\mu$, these deviations only appear at higher orders in the
expansion \eqref{FGTMG}. In the case of even $\mu$, this
nonlinearity turns on a new series of terms in the expansion
\eqref{FGTMG}, although these new terms do not represent new degrees
of freedom.  Note that, even at these higher orders, CDWW still
gives the correct numerical coefficients of the pure derivative
terms.  I expect these features to hold for generic $\mu$.
Table~\ref{table1} shows the first departure from CDWW for $\mu=1$,
2 and 3.

\begin{table}[tbh]
\begin{center}
\begin{tabular}{c|l}
 & first deviation from CDWW \\
\hline $\mu=1$ & $g\indices*{^{(6)}_{++}} = \frac{1}{96}\partial_+^2
    \partial_-^2 F + \frac{1}{8}
    F\partial_-^2 F - \frac{1}{96}
      \left(\partial_-F\right)^2$\\
$\mu=2$ & $g\indices*{^{(8)}_{++}} = \frac{17}{320}
    F\partial\indices*{^2_-}F-\frac{1}{160}\left(\partial\indices{_-}F\right)\indices{^2}$\\
$\mu=3$ & $g\indices*{^{(10)}_{++}} =
    \frac{1}{11520}\partial\indices*{^3_+}\partial\indices*{^3_-}F-\frac{1}{240}\left(\partial\indices{_-}F\right)\indices{^2}+\frac{11}{360}F\partial\indices*{^2_-}F$\\
$\mu=4$ & $g\indices*{^{(12)}_{++}} = \frac{9}{448}
    F\partial\indices*{^2_-}F-\frac{1}{336}\left(\partial\indices{_-}F\right)\indices{^2}$
\end{tabular}
\end{center}\caption{Non-linearities in the $g_{++}$ component.}\label{table1}
\end{table}

The CDWW modes at $\mu=1$ were originally proposed as a
counter-example to the positivity theorem -- while the modes blow up
in the interior, Carlip et.\ al \cite{Carlip2008jk} created
finite-energy superpositions of the modes with negative energy.
However, the original CDWW modes \eqref{CDWWmetric}, as well as
several other proposed counterexamples to positivity, were shown to
develop a linearization instability at second order
\cite{Maloney2009ck,Carlip2009ct}.  For example, the GKP modes
\cite{Giribet2008bw} at second order require the relaxed logarithmic
boundary conditions and thus are not a linear approximation to an
exact solution of chiral gravity.  This is not the case with CDWW --
the nonlinear completion of CDWW, including a background Einstein
metric, interaction terms, and the nonlinear terms of
Table~\ref{table1}, \emph{is} a solution of chiral gravity
satisfying strict Brown-Henneaux boundary conditions. This extended
CDWW should be reconsidered as a potential candidate for violating
the positivity theorem.  The next step, left for the interested
reader, is to construct finite-energy superpositions of this
extended CDWW.

\section{Discussion}
It remains an open question whether physical non-Einstein solutions
of chiral gravity exist.  The technique used here - working from the
boundary in and solving the equations of motion order by order -
offers an alternative approach to the perturbative techniques used
in most of the papers on the subject.  Using this approach, I have
constructed the general solution of TMG with strict Brown-Henneaux
boundary conditions at all values of the mass parameter $\mu\geq1$.
The solutions at each $\mu$ share the same basic structure and can
be written as the sum of an Einstein metric, a purely non-Einstein
metric, and interactions between the two.  The non-Einstein solution
is characterized by a single function $F$ which can be expanded on
to the CDWW modes of \cite{Carlip2008jk}, with nonlinear corrections
to CDWW appearing at higher order.  In particular, the general
solution at the critical value $\mu=1$ contains these extended CDWW
modes, and these modes do not require the relaxed boundary
conditions of log gravity.  Since chiral gravity shares these
features with non-chiral TMG, which is generally thought to be
unstable,\footnote{See \cite{Birmingham2010nb} for an argument that
the BTZ black hole is stable to perturbations at all values of
$\mu$.} this result raises questions about the classical stability
of chiral gravity. However, the task of constructing physically
significant non-Einstein solutions to chiral TMG remains incomplete.
Of the solutions \eqref{chiralsoln} examined in \cite{Compere2010}
all contained either naked singularities or naked closed timelike
curves which violate causality, rendering the solutions unphysical.
However, they considered only a subset of functions $F$ with finite
Fefferman-Graham expansion, and some superposition of CDWW modes
offers an intriguing possibility.

\section{Appendix: FG Expansion of the equations of motion}
This section contains the equations of motion expanded to sixth
order using the standard Fefferman-Graham expansion \eqref{FGEins}.
The full covariant equations are fairly tedious and will not be
include (see \cite{Solodukhin} for the covariant fourth-order
equations).  Instead, the calculations are greatly simplified in the
light-cone gauge using the boundary metric and Levi-Civita
conventions listed in Section 2.  The equations are symmetric under
$\mu \rightarrow -\mu$ and $x^+ \rightarrow x^-$.  Without loss of
generality, we restrict attention to positive $\mu$.  The second
order equations are \cite{Solodukhin}
\begin{eqnarray}
  \text{Tr}\left(g\indices*{_{(0)}^{-1}}g\indices*{_{(2)}}\right)
      &= & -\frac{1}{2}R(g\indices{_{(0)}}) \nonumber \\
  \left(1 - \frac{1}{\mu}\right)\partial_- g\indices*{^{(2)}_{++}}
      &=&0 \nonumber \\
  \left(1 + \frac{1}{\mu}\right) \partial_+ g\indices*{^{(2)}_{--}}
      &=& 0 \label{Firstsoln}
\end{eqnarray}
For all positive $\mu$, $g\indices*{^{(2)}_{--}} = \bar{L}(x^-)$;
however, the equation for $g\indices*{^{(2)}_{++}}$ disappears at
$\mu=1$, leaving this component of the metric unconstrained by the
equations of motion.  At $\mu \neq 1$, the solution reduces to the
Einstein solution $g\indices*{^{(2)}_{++}} = L(x^+)$.

At second order, the $\{r,r\}$ and $\{x^i,x^j\}$ (i,j=1,2) equations
are identically zero, leaving only the $\{r,x^i\}$ equations.  At
higher order, all six equations of motion are present.  However, the
$\{r,x^i\}$ equations are derivatives of the $\{x^i,x^j\}$ equations
and contain no new information. Hence we need only solve the
$\{r,r\}$ and $\{x^i,x^j\}$  equations algebraically in terms of
lower orders. The fourth order equations are
\begin{eqnarray}
  g\indices*{^{(4)}_{+-}} &=& -\frac{1}{4}g\indices*{^{(2)}_{++}}
      g\indices*{^{(2)}_{--}} - \frac{1}{8 \mu}
      \partial\indices*{_-^2}
      g\indices*{^{(2)}_{++}} \nonumber \\
  \left(1-\frac{3}{\mu}\right) g\indices*{^{(4)}_{++}} &=&
      \frac{-1}{4\mu}\partial\indices{_+}\partial\indices{_-}g\indices*{^{(2)}_{++}}
      \nonumber \\
  \left(1 + \frac{3}{\mu}\right) g\indices*{^{(4)}_{--}} &=& 0
  \label{Eom4}
\end{eqnarray}
These equations possess their own critical point $\mu=3$.  When $\mu
\neq 3$, the equations completely determine the components of
$g_{(4)}$ in terms of the $g_{(2)}$, and the solution is either the
chiral solution \eqref{chiralsoln} when $\mu=1$ or the Einstein
solution \eqref{Esoln} for $\mu \neq 1,3$.  However, at $\mu=3$, one
of the equations disappears, leaving $g^{(4)}_{++}$ unconstrained,
and the full solution depends on the three functions
$g^{(2)}_{++}=L(x\indices{^+})$,
$g^{(2)}_{--}=\bar{L}(x\indices{^-})$, and
$g^{(4)}_{++}=F(x\indices{^+},x\indices{^-})$.


This feature is repeated in the sixth order equations:
\begin{eqnarray}
  g\indices*{^{(6)}_{+-}} &=& \left(\frac{-1}{3} -
      \frac{1}{6\mu}\right)g\indices*{^{(2)}_{--}}
      g\indices*{^{(4)}_{++}} - \frac{1}{12\mu}\partial_-^2
      g\indices*{^{(4)}_{++}} \nonumber\\
  \left(1-\frac{5}{\mu}\right) g\indices*{^{(6)}_{++}} &=&  \left(1-
      \frac{6}{\mu}\right)\left[\frac{-2}{3} g\indices*{^{(2)}_{++}}
      g\indices*{^{(4)}_{+-}} - \frac{1}{6}\left(
      g\indices*{^{(2)}_{++}}\right)^2 g\indices*{^{(2)}_{--}}
      \right] + \frac{1}{24}\left(1-\frac{3}{\mu}\right)
      g\indices*{^{(2)}_{++}}\partial_-^2 g\indices*{^{(2)}_{++}}
      \nonumber \\
  & & -\frac{1}{6\mu} \partial_+ \partial_- g\indices*{^{(4)}_{++}}
      + \frac{1}{6\mu} \partial_+^2 g\indices*{^{(4)}_{+-}} +
      \frac{1}{24\mu}g\indices*{^{(2)}_{--}}\partial_=^2
      g\indices*{^{(2)}_{++}} +\frac{1}{24\mu}\left( \partial_-
      g\indices*{^{(2)}_{++}}\right)^2 \nonumber \\
  \left(1+\frac{5}{\mu}\right) g\indices*{^{(6)}_{--}} &=&
      \left(1+\frac{6}{\mu}\right)\left[\frac{-2}{3}
      g\indices*{^{(2)}_{--}}g\indices*{^{(4)}_{+-}} - \frac{1}{6}
      \left(g\indices*{^{(2)}_{--}}\right)^2g\indices*{^{(2)}_{++}}
      \right] + \frac{1}{24}g\indices*{^{(2)}_{--}}\partial_-^2
      g\indices*{^{(2)}_{++}} - \frac{1}{6\mu} \partial_-^2
      g\indices*{^{(4)}_{+-}} \nonumber \\
  & & - \frac{1}{24\mu} g\indices*{^{(2)}_{++}}\partial_-^2
      g\indices*{^{(2)}_{--}} + \frac{1}{6\mu} \partial_- \partial_+
      g\indices*{^{(4)}_{--}} - \frac{1}{24\mu} \left(\partial_-
      g\indices*{^{(2)}_{--}} \right)\left( \partial_-
      g\indices*{^{(2)}_{++}}\right) \label{Eom6}
\end{eqnarray}
The same structure is repeated: at $\mu \neq 5$, $g_{(6)}$ is
determined entirely in terms of the lower order terms.  At the
critical value $\mu=5$, the $g^{(6)}_{++}$ component is
unconstrained.  The equations are easily solved in each of the cases
$\mu=3$, $\mu=5$ and $\mu \neq 1, 3, 5$ and displayed in Table
\eqref{table2}.

\begin{table}[tbh]
\begin{center}
\begin{tabular}{r|c|c|c}
    & $\mu=3$ & $\mu=5$ & $\mu \neq 1,3,5$ \\
\hline $g^{(4)}_{++}$ & $F(x^+,x^-)$ & 0 & 0\\
\hline $g^{(4)}_{+-}$ & $-\frac{1}{4}L(x^+)\bar{L}(x^-)$ &
$-\frac{1}{4}L(x^+)\bar{L}(x^-)$ &
    $-\frac{1}{4}L(x^+)\bar{L}(x^-)$\\
\hline $g^{(6)}_{++}$ & $\frac{1}{12}\partial_+ \partial_- F$ & $F(x^+,x^-)$ & 0\\
\hline $g^{(6)}_{+-}$ & $\frac{-7}{18}F\bar{L} -
\frac{1}{36}\partial_-^2 F$ & $0$ & $0$
\end{tabular}
\end{center}\caption{The non-chiral solutions to sixth order.
Here we use $g^{(2)}_{++}=L(x\indices{^+})$ and
$g^{(2)}_{--}=\bar{L}(x\indices{^-})$. In all cases,
$g^{(4)}_{--}=g^{(6)}_{--}=0$.  Together with \eqref{chiralsoln},
these are the complete set of solutions with the un-modified
Fefferman-Graham expansion \eqref{FGEins} obeying Brown-Henneaux
boundary conditions.}\label{table2}
\end{table}

When $\mu$ does not equal odd integral values, $g_{(6)}$ is
identically zero, and the full solution (with the un-modified
Fefferman-Graham expansion \eqref{FGEins}) is equivalent to the
Einstein solution.  I have expanded the equations out to tenth order
using Maple and have confirmed that $g_{(8)}=g_{(10)}=0$.
Additionally,the Cotton tensor $C_{\mu \nu} =0$ to tenth order.

The solutions at $\mu=3,5$ share several important features.  First,
the full solution is characterized by three functions
$g^{(2)}_{++}=L(x^+)$, $g^{(2)}_{--}=L(x^-)$ and the unconstrained
term $g^{(4)}_{++}=F(x^+,x^-)$ in the case $\mu=3$ or
$g^{(6)}_{++}=F(x^+,x^-)$ in the case $\mu=5$.  The first nonzero
component of the Cotton tensor is
\begin{equation}
  C_{++} = \left\{\begin{array}{cc} \left(\frac{1}{2}\partial\indices{_+}\partial\indices{_-}F\right) e\indices{^{-2r}} & \quad \mu=1\\
  12 F e\indices{^{-2r}} & \quad \mu=3
  \\
  60 F e\indices{^{-4r}} & \quad \mu=5
  \end{array} \right.
\end{equation}
Thus at $\mu=3,5$, the requirement for a non-Einstein solution is
simply that $F\neq 0$. In contrast, the solution at the chiral point
is characterized by only two functions $F(x^+,x^-)$ and
$\bar{L}(x^-)$, and non-Einstein solutions require the more
stringent requirement that $\partial_-F \neq 0$.

The solutions above indicate that the un-modified Fefferman-Graham
expansion \eqref{FGEins} does not capture the propagating modes for
non-odd integral values of the mass parameter.  Instead, the new set
of terms \eqref{FGTMG} is required.  To see this, we can add the
term $e^{(2-m)r}g^{(m)}_{ij}$ to the Fefferman-Graham expansion and
ask what constraints the equations of motion place on $g_{(m)}$.
The equations force the $--$ and $+-$ components to zero, but the
$++$ equation is
\begin{equation}
  \left(1-\frac{m-1}{\mu}\right)g^{(m)}_{++} = 0
\end{equation}
Similar to \eqref{Eom4} and \eqref{Eom6}, this equation disappears
at the critical value $\mu = m-1$, leaving $g^{(m)}_{++}$
unconstrained.  If we call $g^{(m)}_{++} = F(x^+,x^-)$, then the
first non-zero component of the Cotton tensor is $C\indices{_{++}} =
\left(\frac{1}{2}m^3 - \frac{3}{2}m^2+m\right)F
e\indices{^{(2-m)r}}$.

\providecommand{\href}[2]{#2}\begingroup\raggedright\endgroup

\end{document}